\documentclass[aps, pra,
 preprint,
 groupedaddress, amsfonts,
               amsmath, amssymb, showpacs, nofootinbib]{revtex4-1}
\usepackage{microtype}
\usepackage{graphicx}
\usepackage{epstopdf}
\usepackage[utf8]{inputenc}
\usepackage[T1]{fontenc}
\usepackage[usenames,dvipsnames]{xcolor}
\usepackage{hyperref}
\usepackage{amsthm}
\usepackage{tensor}

\usepackage{color}

\begin{document}
\title{Classical versus quantum calculation of radiative electric quadrupole transition rates for hydrogenic states}

\author{Michael Horbatsch}
\affiliation{Department of Physics and Astronomy, York University, Toronto, Ontario M3J 1P3, Canada}

\author{Marko Horbatsch}
\email[]{marko@yorku.ca}
\affiliation{Department of Physics and Astronomy, York University, Toronto, Ontario M3J 1P3, Canada}
\date{\today}
\begin{abstract}
The semiclassical Kepler-Coulomb problem and the quantum-mechanical Schr\"odinger-Coulomb problem are compared
for their predictions of quadrupole E2 transitions. The semiclassical treatment involves an extension of previous work
for the electric dipole transitions (Physical Review A 71, 020501), and rates are derived for $\Delta \ell= 0, \pm 2$ transitions
on the basis of the multipolar properties of the emitted radiation. For the quantum case a derivation is presented within
the Schr\"odinger framework without reference to spin. Comparison of the E2 rates shows reasonable agreement, but not
as good as was found for the electric dipole case.
\end{abstract}
%

%

\maketitle
\section{Introduction}
\label{intro}
The electromagnetic radiation problem for an accelerated charged particle is known in the physics curriculum as the Larmor problem~\cite{LandauLifshitsEM,Jackson}.
The dominant dipolar contribution is treated in many textbooks, as is the associated problem of the radiation reaction force which can be
introduced to account for the transfer of mechanical energy due to the radiating charge.
In the context of quantum mechanics the radiation that is associated with transitions from excited to lower states can be motivated by the
classical radiation problem~\cite{Corney}, but once one introduces matrix elements that connect initial and final states to calculate transition rates
for particular pairs of states the connection to the classical problem appears to be lost or at least deeply hidden.

For the dominant electric dipole transitions from an initial state characterized by principal and orbital angular momentum numbers $(n_i, \ell_i)$
to a final state labeled by $(n_f, \ell_f)$ a reasonably accurate quantum-classical correspondence has been established in Ref.~\cite{PhysRevA.71.020501}.
In this work the semiclassical Kepler-Coulomb trajectories for the initial state were represented through their Fourier-Bessel series,
in principle, as was done originally by Heisenberg and Kramers, just before the advent of modern quantum mechanics.
One modification, however, was the use of semiclassical quantization in the sense of the WKB approximation.
The equidistant frequency spectrum arising from the Fourier-Bessel series was not interpreted literally, but a re-scaling of the contributions from (infinitely) many integer orders
was performed to a non-integer index which took into account the discrete $n$-dependent Balmer spectrum of final available states for the mechanical system.
This re-scaled approach, in which one uses the fact that $n_f=1$ represents the ground state which is not allowed to decay further resulted
in quite accurate transition rates as compared to the Schr\"odinger results, if one made the restriction that $\ell_i>0$.
In fact, it was possible to derive branching ratios associated with angular-momentum-decreasing ($\Delta \ell \equiv \ell_f-\ell_i =-1$) and also angular-momentum-increasing 
($\Delta \ell = +1$) transitions.

In this way it was possible to establish with more confidence than was provided by the general statements in the literature
that using radiated power as predicted by the Larmor expression and dividing by the photon energy, accurate transition rates could be obtained
for electric dipole (E1) allowed transitions. This work was also extended to the case of hydrogenic diamagnetic states~\cite{PhysRevA.72.033405},
for which Hamilton-Jacobi perturbation theory had been developed previously~\cite{PhysRevA.28.7}. More recently this work was extended
to the Stark states~\cite{Horbatsch_2021}. In both cases the Kepler-Coulomb ellipses are perturbed by the external field, yet the classical-quantum correspondence
remains largely intact.

Electric quadrupole (E2) transitions of hydrogenic states (and magnetic dipole M1, in particular) are usually addressed within a relativistic approach
based on solutions to the Dirac equation~\cite{Jitrik_2004}. For our purposes of connecting with the semiclassical Coulomb-Kepler problem
it is, however, sufficient to deal with them at the level of Schr\"odinger-Pauli theory, and if we ignore the magnetic dipole case, we should really
compare to Schr\"odinger solutions, i.e., ignore fine structure. Electric quadrupole transition rates are of interest for 
cases where electric dipole transitions are not allowed, e.g., by a sequence of two such transitions via some intermediate state. 
The E2 selection rules, which are discussed, e.g., in Refs.~\cite{BetheSalpeter,Weissbluth}, then require us to look at the dominant case of $\Delta \ell = -2$,
the case of $\Delta \ell = 0$, and $\Delta \ell = +2$. Interestingly, the calculation of an electric quadrupole transition does come up in an undergraduate
textbook as a problem (Problem 11.31 in Ref.~\cite{GriffithsQM}). Some discussion of how to calculate these transition rates is found in the classic book of Bethe and Salpeter~\cite{BetheSalpeter},
but it comes up short in providing a complete treatment, since it only considers the case of radiation emitted in particular directions. 
Thus, our section on how to calculate these rates in quantum mechanics can fill a gap on its own.

The classical quadrupole radiation problem plays a dominant role in gravitational physics~\cite{PhysRev.131.435}, but the situation is similar to the classical
electromagnetic problem, yet not identical. In the electromagnetic Kepler problem the appearance of opposite charges leads to a time-dependent electric
dipole moment whose acceleration results in E1 Larmor radiation. In the gravitational case where the mass plays the role of charge the center  of mass does
not accelerate in the two-body problem, and thus dipole radiation is not allowed~\cite{LandauLifshits}. Quadrupole radiation can be associated with the third time derivative of the 
quadrupole moment, e.g., as represented by the moment of inertia, and it arises in both Kepler problems. In the Kepler-Coulomb problem it is just a correction
to dipole radiation, in the gravitational Kepler problem it becomes the dominant radiative process.
Motivation for studying the subject of the present paper is of interest in this context, since eventually
a quantum theory of gravity will need to connect quantum and classical gravitational radiation for objects or situations where the classical description may
be an approximation only.

The electric dipole case with its two possible decay paths ($\Delta \ell=\pm1$) was handled by considering
a decomposition of the Kepler motion in such a way that one could identify the parts that correspond to radiation 
carrying away angular momentum of definite sign, but
more detailed work is required to separate the quadrupole transitions into the three contributions 
($\Delta \ell=-2,0,2$).
Thus, one has to extract information from a multipole decomposition of the emitted electromagnetic fields in the radiation zone.
Fortunately, Chapter 9 in Jackson's textbook~\cite{Jackson9} prepares the ground for all required calculations. For recent pedagogical work
beyond the dipole approximation we also refer to Ref.~\cite{Rovenchak}.

Concerning experimental determination of E2 transitions in hydrogenic systems we can make the following remarks.
The measurement of electric quadrupole lifetime in cesium for the $\rm 6S-6D$ and $\rm 6S-5D$ transitions using various methods has been reported in
Refs.~\cite{ZIMMERMANN1974,GLAB1981262,PhysRevA.71.012507} reaching eventually good precision.
Atomic hydrogen spectroscopy~\cite{PhysRevA.93.022513} has reached such high precision that quantum electrodynamics and proton size can be tested, and this
involves spectroscopy which relies on dipole-forbidden transitions (and often two-photon transitions)~\cite{Grinin1061,PhysRevLett.120.183001}.
Sum rules for electric quadrupole transitions in two-electron systems were derived and tested in Ref.~\cite{Bob1974}.
E2 transition rates for atoms with effectively two active electrons in even-parity states were treated non-relativistically in Ref.~\cite{Wilson1991} using 
the formalism provided in Ref.~\cite{PhysRev.57.225} and the book~\cite{CondonOdabasi}, and reasonable agreement was found
with experimental multi-photon ionization rates for neutral silicon atoms~\cite{Ioannidou1990}.
For the alkalis transition rates were calculated within model potential approaches in Refs.~\cite{CELIK2014,epjd28p3},
and expressions for the E2 transition rate for LSJ coupled states can be found there, as well as in Ref.~\cite{CondonHandbook}.
There does not, however, appear to be a direct comparison of experiment versus theory for these transitions in atomic Rydberg levels, and thus,
the present work focuses purely on the comparison of Schr\"odinger and Coulomb-Kepler results.
An interesting recent development, however, is the observation of big changes to normally forbidden transitions in hydrogen when
the atoms are placed in a nano-scale environment with external fields~\cite{KimKim2018}.

The paper is organized as follows. In Sect.~\ref{sec:theory1} the semiclassical E2 rates for $(n_i,\ell_i)\to (n_f,\ell_f)$ transitions
are derived from general expression given by Jackson~\cite{Jackson9}.
In Sect.~\ref{sec:theory2} we provide the corresponding calculation for the quantum-mechanical E2 transition rates.
In Sect.~\ref{sec:res} we provide some comparisons for a few cases, indicating strengths and weaknesses of the semiclassical approach.
We present graphical results for transitions from $(n_i=30,\ell_i=15)$ to illustrate the decay possibilities for $\Delta \ell =-2,0,2$, in terms of their
branching ratios as a function of emitted photon frequency, and show how the semiclassical results compare with the quantum ones. 
This is followed up by a table which compares absolute transition rates for a number of cases.
The paper ends with concluding remarks in Sect.~\ref{sec:conclusions}.

\section{Theory}
\label{sec:model}

\subsection{Semiclassical approach}
\label{sec:theory1}
Here we analyze the electric quadrupole radiation emitted by a classical model of a hydrogenic atom, i.e.
a point particle of charge $-e$ moving along a Kepler 
ellipse. The theory of the multipole expansion of radiation fields is developed in Ref.~\cite{Jackson9}. 
From Jackson's equation (9.155), the radiated power from the $\ell = 2$ electric multipole is given by
\begin{equation}
\label{eq:jackson_power}
P = \int_{0}^{\infty} d\omega \frac{c}{2\epsilon_{0}\omega^2} \sum_{m=-2}^{2}\left|a_{E}^{(\omega)}(2,m)\right|^2 \,.
\end{equation}
Note that the discussion in Ref.~\cite{Jackson9} considers only a single Fourier mode, whereas here we need multiple
Fourier modes to describe ellipses with non-zero eccentricity, hence the integral over $\omega$.

The electric quadrupole coefficients $a_{E}^{(\omega)}(2,m)$ appearing in (\ref{eq:jackson_power}) may be computed using  equations 
(9.169)-(9.170) from Ref.~\cite{Jackson9}:
\begin{equation}
\label{eq:elquad_coef}
a_{E}^{(\omega)}(2,m) = \frac{\omega^4}{15i c^3}\sqrt{\frac{3}{2}} \int |\vec{x}|^2 \, Y_{2,m}^{*} \,
\rho_{\omega} \, d^{3}\vec{x} \,,
\end{equation}
where the relevant spherical harmonics $Y_{\ell,m}$ are given explicitly on pages 108-109 of Ref.~\cite{Jackson}, and
\begin{equation}
\rho_{\omega}(\vec{x}) = \frac{1}{\pi}\int_{-\infty}^{\infty} \rho(\vec{x},t)e^{i \omega t} dt 
\end{equation}
are the Fourier components of the electric charge density $\rho(\vec{x},t)$. For a point particle with 
charge $-e$ and trajectory $\vec{r}(t)$, we have
\begin{equation}
\rho(\vec{x},t) = -e \delta^{(3)}(\vec{x} - \vec{r}(t)) \,.
\end{equation}
In principle, in the classical model of an atom,
the above equation should contain an additional term describing the presence of the nucleus, but we do not write it down,
because it does not directly contribute to the radiation as long as reduced-mass effects are neglected. 

In order to evaluate the electric quadrupole coefficients (\ref{eq:elquad_coef}), we need an explicit description of the
Kepler trajectory, which is known as a Fourier series:
\begin{eqnarray}
\vec{r}(t) &=& \Bigl( x(t) , y(t) , 0 \Bigr) \,,
\\
\left[ x(t) \right]^{2} &=& a^2 \sum_{k=0}^{\infty}{A}_{k} \cos(k \Omega t) \,,
\\
\left[ y(t) \right]^{2} &=& a^2 \sum_{k=0}^{\infty}{B}_{k} \cos(k \Omega t) \,,
\\
x(t) \cdot y(t) &=& a^2 \sum_{k=1}^{\infty}{C}_{k} \sin(k \Omega t) \,,
\end{eqnarray}
where $a$ and $\Omega$
are the semimajor axis and angular frequency of the Kepler orbit, respectively, and 
\begin{eqnarray}
\label{eq:fourcoefA}
{A}_{k} &=& \frac{1}{k} \left[
J_{k-2}(k\epsilon) - J_{k+2}(k\epsilon) -2\epsilon (J_{k-1}(k\epsilon) -J_{k+1}(k\epsilon))
\right] \,,
\\
\label{eq:fourcoefB}
{B}_{k} &=& \frac{1-\epsilon^2}{k} \left[
J_{k+2}(k\epsilon) -J_{k-2}(k\epsilon)
\right] \,,
\\
\label{eq:fourcoefC}
{C}_{k} &=& \frac{\sqrt{1-\epsilon^2}}{k} \left[
J_{k+2}(k\epsilon) + J_{k-2}(k\epsilon) -\epsilon ( J_{k+1}(k\epsilon) + J_{k-1}(k\epsilon)) 
\right] \,,
\\
\label{eq:fourcoefA0}
{A}_{0} &=& \frac{1}{2}(1+4\epsilon^2) \,,
\\
\label{eq:fourcoefB0}
{B}_{0} &=& \frac{1}{2}(1-\epsilon^2) \,,
\end{eqnarray}
are the Fourier coefficients~\cite{Maggiore}.
The standard notation $J_{k}(z)$ is used for the Bessel function of order $k$~\cite{Abramowitz}.
The constant terms ${A}_{0}$ and ${B}_{0}$ do not enter into any final radiation formulas, and are
only provided for completeness. The quantity $\epsilon$ 
is the eccentricity of the Kepler ellipse.
By means of semiclassical quantization~\cite{PhysRev.51.669}, properties of the Kepler orbit may be related to
the principal quantum number $n$, and the orbital quantum number $\ell$:
\begin{equation}
a = \frac{n^2\hbar}{Z\alpha m_{e}c} \,,
\qquad
\Omega = \frac{Z^2 \alpha^2 m_{e} c^2}{n^3 \hbar} \,,
\qquad
\epsilon = \sqrt{1 - \frac{(\ell + \tfrac{1}{2})^2}{n^2}} \,.
\end{equation}
Here $Z$ is the atomic number (so the nucleus has charge $+Ze$), $m_{e}$ is the electron mass, 
and $\alpha = e^2/(4\pi\epsilon_{0}\hbar c) \sim 1/137$ is the fine-structure constant. 

With the Fourier series in hand, equation (\ref{eq:elquad_coef}) may now be explicitly evaluated, yielding
\begin{eqnarray}
\label{eq:acoef0}
a_{E}^{(\omega)}(2,0) &=& -\frac{iea^2\omega^4}{8c^3} \sqrt{\frac{2}{15\pi}}
\sum_{k=1}^{\infty}({A}_{k} + {B}_{k})\delta(\omega - k\Omega) \,,
\\
\label{eq:acoef2}
a_{E}^{(\omega)}(2,\pm 2) &=& \frac{iea^2\omega^4}{8\sqrt{5\pi}c^3} 
\sum_{k=1}^{\infty}({A}_{k} - {B}_{k} \pm 2{C}_{k})\delta(\omega - k\Omega) \,,
\\
\label{eq:acoef1}
a_{E}^{(\omega)}(2,\pm 1) &=& 0 \,.
\end{eqnarray}
In the above equations, zero-frequency terms involving $\delta(\omega)$ have been dropped, since
they do not contribute to the radiated power (\ref{eq:jackson_power}). Also, the vanishing of the 
$m=\pm 1$ components is a consequence of choosing the coordinate system such that the Kepler ellipse
lies in the $xy$ plane.

Substituting equations (\ref{eq:acoef0})-(\ref{eq:acoef1}) into the power formula (\ref{eq:jackson_power}), 
we obtain
\begin{eqnarray}
\label{eq:power_m0}
P_{m=0} &=& \frac{e^2 a^4 \Omega^6}{960\pi\epsilon_{0}c^5}\sum_{k=1}^{\infty}k^6({A}_{k} + {B}_{k})^2 \,,
\\
\label{eq:power_mp2}
P_{m=+2} &=& \frac{e^2 a^4 \Omega^6}{640\pi\epsilon_{0}c^5}\sum_{k=1}^{\infty}k^6({A}_{k} - {B}_{k} + 2 {C}_{k})^2 \,,
\\
\label{eq:power_mm2}
P_{m=-2} &=& \frac{e^2 a^4 \Omega^6}{640\pi\epsilon_{0}c^5}\sum_{k=1}^{\infty}k^6({A}_{k} - {B}_{k} - 2 {C}_{k})^2 \,.
\end{eqnarray}
From the calculation and discussion in sections 9.8 and 9.11 of Ref.~\cite{Jackson9} it follows that:
\begin{itemize}
\item 
The $k^{\rm th}$ term in 
equation (\ref{eq:power_m0}), divided by $\hbar k \Omega$, is the rate for $\Delta \ell = 0$ transitions.
\item 
The $k^{\rm th}$ term in 
equation (\ref{eq:power_mp2}), divided by $\hbar k \Omega$, is the rate for $\Delta \ell = -2$ transitions.
\item 
The $k^{\rm th}$ term in 
equation (\ref{eq:power_mm2}), divided by $\hbar k \Omega$, is the rate for $\Delta \ell = +2$ transitions.
\end{itemize}
Explicitly, and in terms of quantities familiar to atomic physicists, the classical rates (before harmonic index rescaling) 
are given by 
\begin{equation}
R_{k,\Delta \ell} = \frac{Z^6 \alpha^7 m_{e} c^2}{60 n^7 \hbar} \tilde{R}_{k,\Delta \ell} \,,
\label{eq:rateSCL}
\end{equation}
with
\begin{eqnarray}
\tilde{R}_{k,\Delta \ell = 0} = \frac{k^5}{4}({A}_{k} + {B}_{k})^2 \,,
\\
\tilde{R}_{k,\Delta \ell = \pm 2} = \frac{3k^5}{8}({A}_{k} -{B}_{k} \mp 2{C}_{k})^2 \,.
\label{eq:ratedlpm}
\end{eqnarray}
These expressions complement the previously obtained semiclassical rates for E1 transitions~\cite{PhysRevA.71.020501}:
\begin{equation}
R_{k,\Delta \ell =\pm 1} = \frac{2 Z^4 \alpha^5 m_{e} c^2}{3 n^5 \hbar} k \left( J_k'(k \epsilon) \mp \sqrt{\epsilon^{-2}-1} J_k(k \epsilon) \right)^2  \,.
\label{eq:rateE1}
\end{equation}

The given rates for E1 or E2 transitions are in terms of Fourier order $k$, i.e., correspond to an equidistant photon energy spectrum $\hbar k \Omega$.
In order to compare with transitions to allowed hydrogenic states with energies $E_{n'}$ in accord with the Balmer formula they can be rescaled to non-integer order $k_{\Delta n}$
and weighted appropriately, as explained in Ref.~\cite{PhysRevA.71.020501}, and in Section~\ref{sec:res} (cf. eq.~(\ref{eq:resc}) and Fig.~\ref{fig:Fig1}).
While the Fourier rates depend only on the properties of the initial state $(n, \ell )$ whose spontaneous decay is being considered, the rescaling introduces
dependence on the final state, and makes comparison with quantum transition rates to states $(n', \ell' )$ possible.

\subsection{Quantum mechanical approach}
\label{sec:theory2}

As mentioned in the introduction most publications reporting electric quadrupole transition rates quote formulae given
in handbooks for the case of total angular momentum eigenstates which include spin-orbit coupling. These can be derived
using the formalism of the Wigner-Eckart theorem~\cite{Weissbluth,Shore}. Given that we are interested in quantum-classical correspondence
we would like to work only with orbital angular momentum eigenstates. The calculation can be carried out directly, and this approach
will be followed here.

Following Ref.~\cite{GriffithsQM} (p. 556, eq.~ 11.130, and including an apparently omitted factor of $1/4$), the quadrupole
transition rate is given by 
\begin{equation}
\label{eq:rat_quad}
R_{n\ell m \to n' \ell' m'} = \frac{e^2 \omega^5}{4 \pi \epsilon_{0} \hbar c^5} \,
\left | \sum_{i,j}{\epsilon_{i}k_{j} \langle n'\ell' m'|r_{i}r_{j}|n \ell m \rangle } \right |^2 \,.
\end{equation}
Here $r_{i}$ are the components of the electron position vector $\vec{r}$, $\omega$ is the circular photon frequency,
$\epsilon_{i}$ are the components of the (unit) photon polarization vector $\hat{\epsilon}$, 
and $k_{i}$ are the components of the (unit) photon propagation direction vector $\hat{k}$. The latter two
vectors must be orthogonal, i.e.
\begin{equation}
\label{eq:dir_pol_constr}
\hat{k} \cdot \hat{\epsilon} = 0 \,.
\end{equation}
The first thing that we need to do, is average 
the quadrupole rate (\ref{eq:rat_quad}) over all photon directions $\hat{k}$ and polarizations $\hat{\epsilon}$, 
while enforcing the constraint (\ref{eq:dir_pol_constr}). To do this, begin by introducing polar and azimuthal
angles for $\hat{\epsilon}$:
\begin{equation}
\epsilon_{x} = \sin\theta \cos \phi \,,
\qquad
\epsilon_{y} = \sin\theta \sin \phi \,,
\qquad
\epsilon_{z} = \cos \theta \,.
\end{equation}
For fixed values of $\theta$ and $\phi$, a geometrical argument involving rotation matrices
can be used to show that the vector $\hat{k}$ 
is constrained to lie on a unit circle that may be parametrized according to
\begin{eqnarray}
k_{x} &=& \cos \phi \cos \theta \cos \psi - \sin \phi \sin \psi \,,
\\
k_{y} &=& \sin\phi \cos \theta \cos \psi + \cos\phi \sin \psi \,,
\\
k_{z} &=& -\sin\theta \cos \psi \,,
\end{eqnarray}
where $0 \le \psi < 2\pi$.
The direction- and polarization-averaged rate is now given by
\begin{equation}
\label{eq:rat_quad_avg}
R_{n\ell m \to n' \ell' m'} = \frac{e^2 \omega^5}{4\pi \epsilon_{0} \hbar c^5}
\, \frac{1}{4\pi} \int_{0}^{\pi}d\theta \int_{0}^{2\pi} \sin \theta d\phi \,
\frac{1}{2\pi} \int_{0}^{2\pi} d\psi \, 
\Bigl |\sum_{i,j}{\epsilon_{i}k_{j} Q_{ij}} \Bigr |^2 \,,
\end{equation}
where we have introduced the notation
\begin{equation}
Q_{ij} = Q_{ji} = \langle n'\ell' m'|r_{i}r_{j}|n \ell m \rangle
\end{equation}
for the quadrupole matrix elements. Evaluating the angular integrals using a computer algebra program
we obtain
\begin{eqnarray}
R_{n\ell m \to n' \ell' m'} &=& \frac{e^2 \omega^5}{4\pi \epsilon_{0} \hbar c^5} \Biggl[
\frac{1}{15} \left( |Q_{xx}|^2 + |Q_{yy}|^2 + |Q_{zz}|^2 \right)
\nonumber
\\
&&
\phantom{ \frac{e^2 \omega^5}{\pi \epsilon_{0} \hbar c^5} \Biggl[ }
+ \frac{1}{5} \left( |Q_{xy}|^2 + |Q_{xz}|^2 + |Q_{yz}|^2 \right)
\nonumber
\\
&&
\phantom{ \frac{e^2 \omega^5}{\pi \epsilon_{0} \hbar c^5} \Biggl[ }
-\frac{1}{30} \left( Q_{xx}Q^{*}_{yy} + Q_{xx}^{*}Q_{yy} \right)
\nonumber
\\
&&
\phantom{ \frac{e^2 \omega^5}{\pi \epsilon_{0} \hbar c^5} \Biggl[ }
-\frac{1}{30} \left( Q_{xx}Q^{*}_{zz} + Q_{xx}^{*}Q_{zz} \right)
\nonumber
\\
&&
\phantom{ \frac{e^2 \omega^5}{\pi \epsilon_{0} \hbar c^5} \Biggl[ }
-\frac{1}{30} \left( Q_{yy}Q^{*}_{zz} + Q_{yy}^{*}Q_{zz} \right)
\Biggr] \,.
\label{eq:Rnlm}
\end{eqnarray}
Now we compute the six quadrupole matrix elements. Using explicit formulas for the 
spherical harmonics $Y_{\ell, m}$ with $\ell = 2$, we obtain
\begin{eqnarray}
Q_{xx} &=&  \tensor*{I}{*^{n'}_{\ell'}^{n}_{\ell}} \left\{
\sqrt{\frac{2\pi}{15}} 
\left[
\tensor*{\Lambda}{*^{\ell'}_{m'}^{2}_{2}^{\ell}_{m}}
+ \tensor*{\Lambda}{*^{\ell'}_{m'}^{2}_{-2}^{\ell}_{m}}
\right]
- \frac{1}{3}
\sqrt{\frac{4\pi}{5}} \tensor*{\Lambda}{*^{\ell'}_{m'}^{2}_{0}^{\ell}_{m}}
+ \frac{1}{3} \delta_{\ell \ell'} \delta_{m m'}
\right\} \,,
\\
Q_{yy} &=&  \tensor*{I}{*^{n'}_{\ell'}^{n}_{\ell}} \left\{
-\sqrt{\frac{2\pi}{15}} 
\left[
\tensor*{\Lambda}{*^{\ell'}_{m'}^{2}_{2}^{\ell}_{m}}
+ \tensor*{\Lambda}{*^{\ell'}_{m'}^{2}_{-2}^{\ell}_{m}}
\right]
- \frac{1}{3}
\sqrt{\frac{4\pi}{5}} \tensor*{\Lambda}{*^{\ell'}_{m'}^{2}_{0}^{\ell}_{m}}
+ \frac{1}{3} \delta_{\ell \ell'} \delta_{m m'}
\right\} \,,
\\
Q_{zz} &=&  \tensor*{I}{*^{n'}_{\ell'}^{n}_{\ell}} \left\{
\frac{2}{3}
\sqrt{\frac{4\pi}{5}} \tensor*{\Lambda}{*^{\ell'}_{m'}^{2}_{0}^{\ell}_{m}}
+ \frac{1}{3} \delta_{\ell \ell'} \delta_{m m'}
\right\} \,,
\\
Q_{xy} &=&  \tensor*{I}{*^{n'}_{\ell'}^{n}_{\ell}} \left\{
\frac{1}{i}
\sqrt{\frac{2\pi}{15}} 
\left[
\tensor*{\Lambda}{*^{\ell'}_{m'}^{2}_{2}^{\ell}_{m}}
- \tensor*{\Lambda}{*^{\ell'}_{m'}^{2}_{-2}^{\ell}_{m}}
\right]
\right\} \,,
\\
Q_{xz} &=&  \tensor*{I}{*^{n'}_{\ell'}^{n}_{\ell}} \left\{
\frac{1}{2}
\sqrt{\frac{8\pi}{15}} 
\left[
\tensor*{\Lambda}{*^{\ell'}_{m'}^{2}_{-1}^{\ell}_{m}}
- \tensor*{\Lambda}{*^{\ell'}_{m'}^{2}_{1}^{\ell}_{m}}
\right]
\right\} \,,
\\
Q_{yz} &=&  \tensor*{I}{*^{n'}_{\ell'}^{n}_{\ell}} \left\{
\frac{i}{2}
\sqrt{\frac{8\pi}{15}} 
\left[
\tensor*{\Lambda}{*^{\ell'}_{m'}^{2}_{-1}^{\ell}_{m}}
+ \tensor*{\Lambda}{*^{\ell'}_{m'}^{2}_{1}^{\ell}_{m}}
\right]
\right\} \,.
\end{eqnarray}
Here, we have introduced the notation
\begin{equation}
\tensor*{I}{*^{n'}_{\ell'}^{n}_{\ell}} = 
\int_{0}^{\infty} dr r^4 R^{*}_{n' \ell'}(r) R_{n \ell}(r) 
\label{eq:radint}
\end{equation}
for the radial integrals ($R_{n \ell}(r)$ are the real-valued hydrogenic radial wavefunctions), and 
\begin{equation}
\tensor*{\Lambda}{*^{\ell'}_{m'}^{\ell_{1}}_{m_{1}}^{\ell_{2}}_{m_{2}}} = 
\int_{0}^{\pi} d\theta \int_{0}^{2\pi} \sin \theta d\phi \, 
Y^{*}_{\ell', m'}(\theta,\phi) Y_{\ell_{1}, m_{1}}(\theta,\phi) Y_{\ell_{2}, m_{2}}(\theta,\phi)
\end{equation}
for the angular integrals. These angular integrals may be expressed in terms of 
Clebsch-Gordan coefficients, see, e.g., Ref.~\cite{Sakurai} (p. 218, eq. 3.393):
\begin{equation}
\tensor*{\Lambda}{*^{\ell'}_{m'}^{\ell_{1}}_{m_{1}}^{\ell_{2}}_{m_{2}}}
=
\sqrt{\frac{(2\ell_{1}+1)(2\ell_{2}+1)}{4\pi (2\ell' + 1)}}
\tensor*{C}{*^{\ell_{1}}_{0}^{\ell_{2}}_{0}^{\ell'}_{0}}
\tensor*{C}{*^{\ell_{1}}_{m_{1}}^{\ell_{2}}_{m_{2}}^{\ell'}_{m'}} \,.
\end{equation}

To proceed further we observe some properties of the transition rate~(\ref{eq:Rnlm}):
summing over the magnetic quantum numbers $m'$ for the final state we find numerically 
an answer that is independent of the initial-state quantum number $m$.
Usually the initial $m$-value is not specified, and one averages over $m$, but needs to sum over
all final $m'$. This leads to the idea of defining
the strength for $(n, \ell) \to (n', \ell')$ E2 transitions,
in analogy to the detailed discussion for electric dipole transitions in Ref.~\cite{Weissbluth} (chapter 23):
\begin{equation}
S_{n \ell \to n' \ell'}=\sum_{m,m'}{\tilde R_{n \ell m \to n' \ell' m'}} \ ,
\label{eq:Snl}
\end{equation}
where $\tilde R_{n \ell m \to n' \ell' m'}$ is given by~(\ref{eq:Rnlm}) with the prefactor omitted.

Using angular momentum algebra it is possible to simplify the result such that 
\begin{equation}
S_{n \ell \to n' \ell'}=\frac{(2\ell+1)(2\ell'+1)}{15} 
\begin{pmatrix}
\ell' & 2  & \ell \\
0    & 0  & 0
\end{pmatrix}^2
(\tensor*{I}{*^{n'}_{\ell'}^{n}_{\ell}})^2
 \ ,
\label{eq:Snl2}
\end{equation}
where the term in round brackets is a Wigner 3j symbol and the squares of the radial integrals~(\ref{eq:radint}) need to be computed.
In order to compare with the semiclassical transition rates one needs to include the pre-factor given in~(\ref{eq:Rnlm}), and divide by the 
multiplicity of the initial state in order to account for statistical averaging.

In order to compare with literature values obtained within relativistic quantum mechanics~\cite{Jitrik_2004} 
one has to convert the transition strength to states with defined total angular momentum
in accord with the LSJ coupling scheme. This can be accomplished as outlined by Weissbluth~\cite{Weissbluth} for the dipole case (eq. 23.1-25),
but using the Wigner-Eckart theorem (eq. 6.3-39) for the tensor rank $q=2$. This results in the transition strength expression
\begin{equation}
S_{n\ell j\to n' \ell' j'}= (2j+1)(2j'+1)\begin{Bmatrix}
j' & 2  & j \\
l    & \tfrac{1}{2}  & l'
\end{Bmatrix}^2
S_{n\ell\to n' \ell'} \ ,
\label{eq:Snlj}
\end{equation}
where generally $j=\ell \pm \tfrac{1}{2}$, and the term in curly brackets is a Wigner 6j symbol.

The final expressions then become the following: for comparison with the semiclassical rates we have
\begin{equation}
R_{n\ell\to n' \ell'}=  \frac{e^2 \omega^5}{4\pi \epsilon_{0} \hbar c^5} \frac{1}{2 \ell+1}  S_{n\ell\to n' \ell'} \ ,
\label{eq:Rnlfin}
\end{equation}
while for comparison with the relativistic calculations we use
\begin{equation}
R_{n\ell j\to n' \ell' j'}= \frac{e^2 \omega^5}{4\pi \epsilon_{0} \hbar c^5} \frac{1}{2 j+1} S_{n\ell j\to n' \ell' j'} \ .
\label{eq:Rnljfin}
\end{equation}

\section{Results}
\label{sec:res}

We consider results for hydrogen only, i.e., $Z=1$. We have checked that the quantum  LSJ coupling results based on the Schr\"odinger equation treatment, i.e.,
eq.~(\ref{eq:Snlj})
agrees to several digits with the Dirac equation based results from Ref.~\cite{Jitrik_2004}, which are available in tabular form for many states~\cite{Bunge}.
For higher values of $Z$ relativistic effects due to the energy differences between initial and final states, as well as the character of the radial functions will
become important. Thus, our comparison between semiclassical and quantum results will be based solely on the $(n,\ell) \to (n',\ell')$ transitions.
Electric quadrupole E2 transitions are stated in the literature to be five orders of magnitude weaker than E1 transitions,
which can be estimated from the prefactors given, e.g., in eqs.~(\ref{eq:rateSCL}) vs.~(\ref{eq:rateE1}).
Therefore, the most important comparisons will be for those cases where a dipole transition cascade will not connect two levels. It would be an interesting
exercise in its own right to check how a sequence of two E1 transitions (e.g., when $n-n' \ge 2$) would compete against a direct E2 transition.

We consider first a high-$n$, intermediate-$\ell$ initial state to allow for pathways involving $\Delta \ell = -2, 0, 2$ transitions, and begin with a comparison of the overall 
transition rates to lower principal quantum number $n'=n-\Delta n$ ($\Delta n$ is positive). In the classical limit electromagnetic radiation at Fourier
order $k$ is approximately treated as corresponding to a transition to a lower state according to $\Delta n =k$. 
The Balmer spectrum is not equidistant, however, and therefore the re-scaling
approach of Ref.~\cite{PhysRevA.71.020501} was introduced to re-scale the Fourier coefficients to non-integer order
\begin{equation}
k_{\Delta n} = \frac{n}{2}\left[( 1-\Delta n /n)^{-2} -1\right] 
\label{eq:resc}
\end{equation}
which also implies a weighting of this effective harmonic transition rate by a factor of $( 1-\Delta n /n)^{-3}$, (cf. Eqs.~(22,23) in Ref.~\cite{PhysRevA.71.020501}).

\begin{figure}
\begin{center}$
\begin{array}{cc}
\resizebox{0.5\textwidth}{!}{\includegraphics{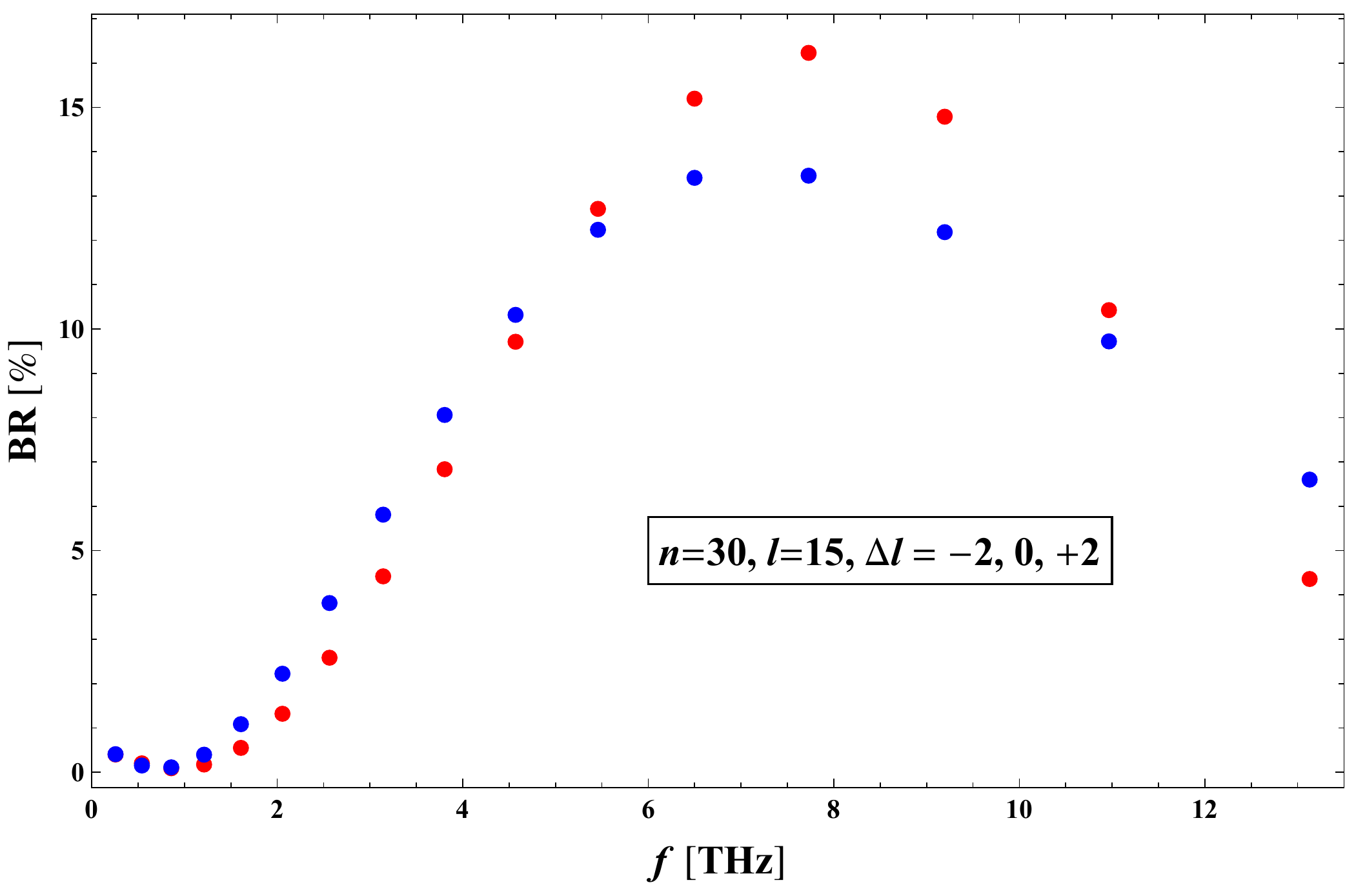}}&
\resizebox{0.5\textwidth}{!}{\includegraphics{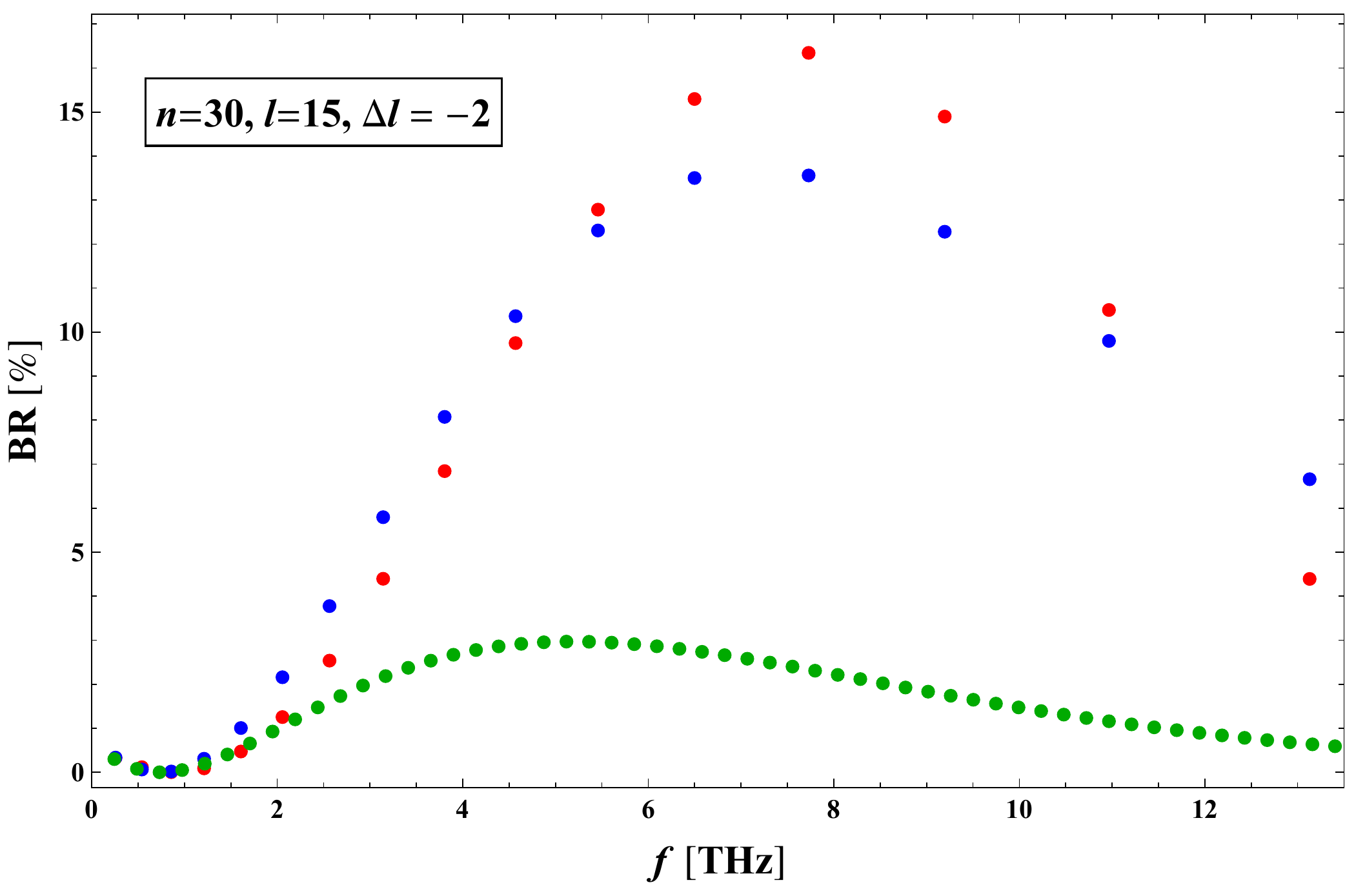}}\\
\resizebox{0.5\textwidth}{!}{\includegraphics{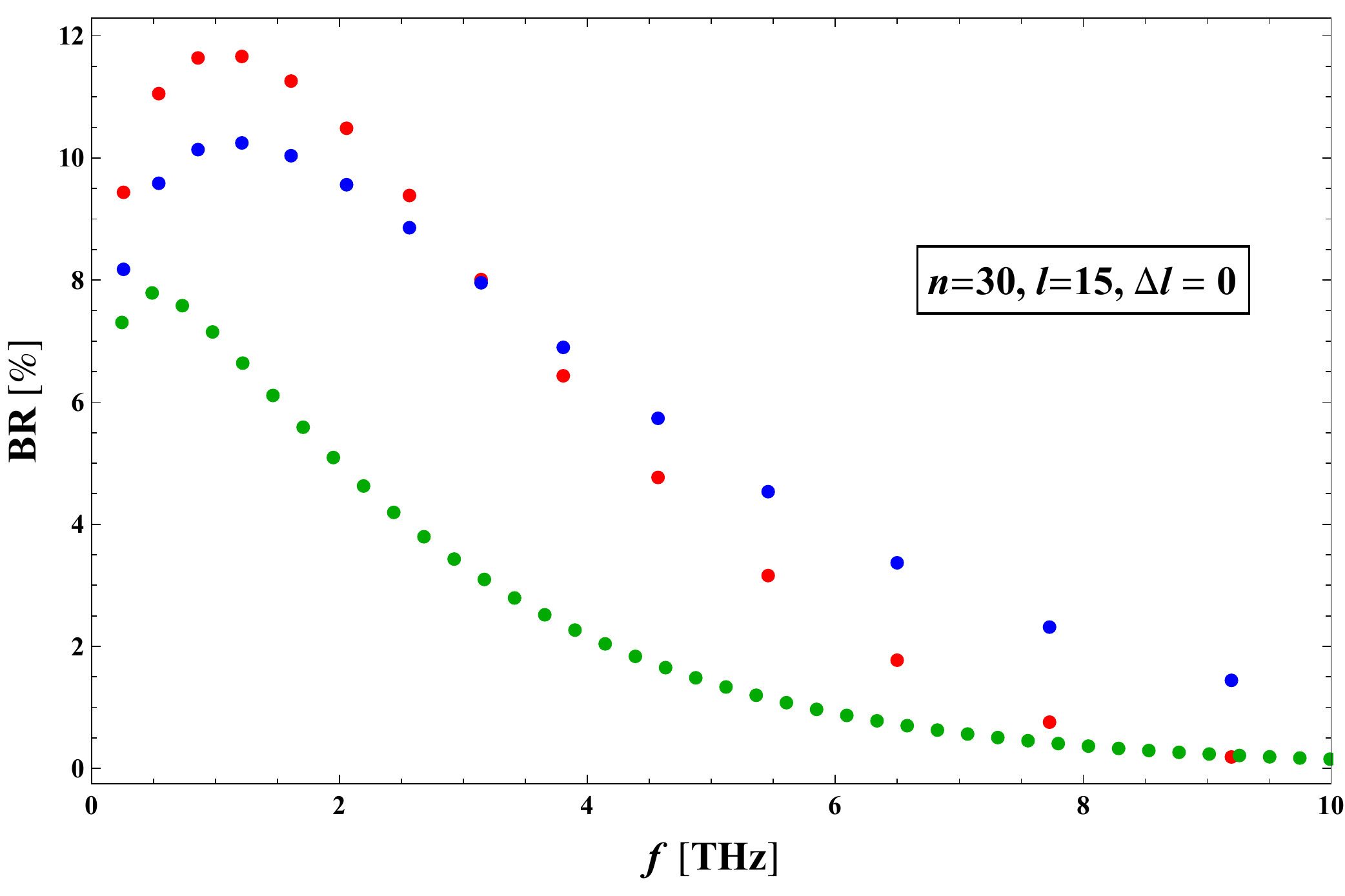}}&
\resizebox{0.5\textwidth}{!}{\includegraphics{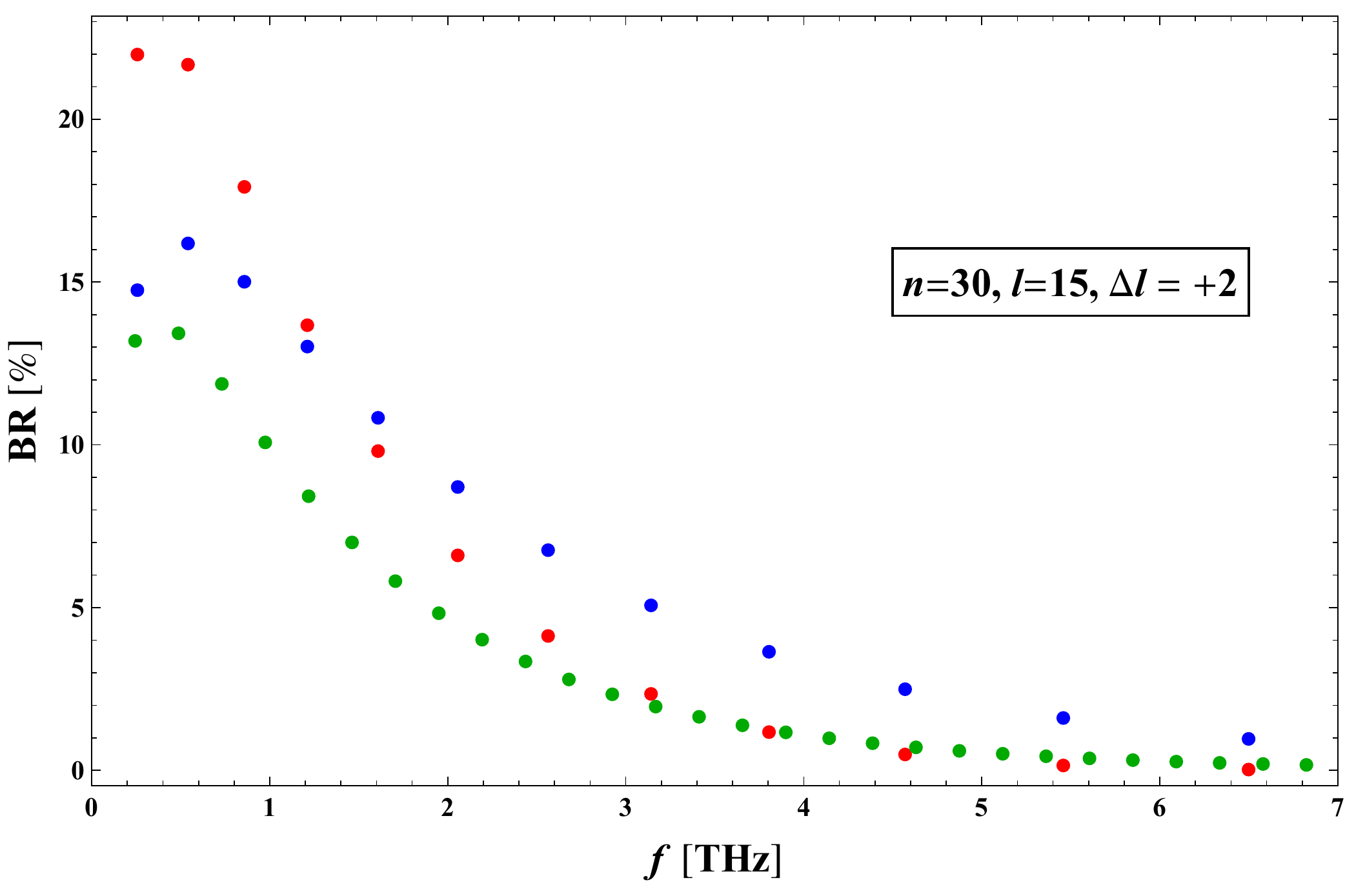}}
\end{array}$
\caption{%
Branching ratios for E2 quadrupole transitions from the state $(n=30, \ell=15)$ to states $n'=n-\Delta n$ as a function of emitted radiation frequency $f$.
The red dots represent quantum-mechanical ratios, the green dots show the classical ratios, and the blue dots show the re-scaled
classical branching ratios according to eq.~(\ref{eq:resc}).
The top left panel shows the quantum and re-scaled classical branching ratios as a function of photon frequency only, i.e., irrespective of $\Delta \ell$, 
while the other three panels show the $(\Delta n, \Delta \ell)$-dependent
ratios separately for $\Delta \ell =-2$,  $\Delta \ell =0$, and
 $\Delta \ell =+2$ transitions.
}
\label{fig:Fig1}
\end{center}
\end{figure}

The top left panel of Fig.~\ref{fig:Fig1} shows the $(\Delta n)$-dependent branching fractions in percent irrespective of whether angular momentum is decreasing, remains the same, or is
increasing for the transitions. The points correspond to transitions with $\Delta n = 1,2,...$ from left to right, and are only shown for the quantum and re-scaled classical
results. A peak occurs at $\Delta n =13$, or about $6-8 \ \rm THz$ in frequency.
At this point the rates exceed the low-$\Delta n$ results by almost two orders of magnitude. 
The minimum at $\Delta n=3$ which is obtained both in the quantum and in the re-scaled
classical calculation can be explained when looking at the $\Delta \ell=-2$ transitions alone.

Considering now the branching fractions associated with angular momentum decreasing transitions (top right panel) we observe a number of things.
First of all, the green points show the classical Fourier spectrum of frequencies, which is equidistant (harmonic). At small $k \approx \Delta n$ it follows the
quantum results quite well, also exhibiting a minimum.
There is a near-cancellation
at this Fourier order, i.e., at $k=3$. This is happening in the combination of Fourier coefficients in eq.~(\ref{eq:ratedlpm}), caused by
the fact that there is a sign change in the combination $({A}_{k} -{B}_{k} + 2{C}_{k})$ for some (non-integer) value of $k=k_c$.
In the quantum calculation it is the expectation value $\langle n, \ell |r^2 | n', \ell-2 \rangle$ that changes sign between two subsequent values of $n'=n-\Delta n$.
The result is that for given $(n, \ell)$ at particular $\Delta n$ a minimum occurs. For higher values of $n$ this minimum is shifted to higher $\Delta n$,
or classical Fourier order $k$.

With increasing $\Delta n$ the quantum transition energies spread apart according to the Balmer formula, thus causing a distinctive pattern in the
frequency spectrum. The transition rates increase with $\Delta n$ due to the phase space factor, i.e., higher photon energies are favored.
Eventually, however, there is a turnover, because the radial matrix elements become unfavorable. The classical Fourier series (green points) 
also displays a maximum (at around 5 THz). Many equally-spaced Fourier orders contribute, and thus the branching fractions are smaller.
They need to be added up by a binning procedure to take into account the allowed final states. This is accomplished in a continuous way
by the re-scaling given in eq.~(\ref{eq:resc}). The re-scaling makes use of non-integer index $k_{\Delta n}$, and a weight is applied to
the result. The blue points show that this is effective in the sense that a similar distribution is obtained as for the quantum case.

While the quantum calculation depends on the initial- and final-state wavefunctions, the re-scaled semiclassical calculation makes use of the initial-state
Fourier series, but then applies the Balmer formula to sum up the contributions from multiple Fourier orders that correspond to a transition
to final $n'=n-\Delta n$.
At moderate values of $\Delta n$ there is an overestimation of the transition rates (blue points), but at high $\Delta n$ a very similar maximum as for the 
quantum results
can be observed, and then a cross-over, where the highest-allowed frequency contribution is overestimated.

For the $\Delta \ell=0$ transitions shown in the bottom left panel no particularly interesting feature arises, the transition rates
rise to form a maximum at $\Delta n = 3-4$, and then fall off. The classical Fourier rates (green points) when binned effectively
to correspond to quantized final-state energies using eq.~(\ref{eq:resc}) underestimate the quantum rates at the maximum,
and then overestimate them strongly at the highest allowed frequencies.
A similar behavior is found for the $\Delta \ell = +2$ transitions shown in the bottom right panel. 

The decays from a high-$n$, intermediate-$\ell$ initial state discussed above allowed us to contrast the classical Fourier radiation spectrum
from the elliptic orbit that represents the initial state with the available space of quantum levels. Through an effective continuous binning process
(eq.~(\ref{eq:resc})) we demonstrated how quantum transitions can be made compatible with a semiclassical picture. For the case shown it works reasonably
well for the total $n \to n'$ transition rates at not too large values of $\Delta n = n-n'$. It is based mostly on a reasonable treatment of $\Delta \ell = -2$ transitions,
and a moderately successful representation of the case of $\Delta \ell = 0$, while the overall less important case of $\Delta \ell = +2$ was seen to
be described more poorly in the semiclassical analysis.

In order to show results for cases where the absolute transition rates are larger we present
in table~\ref{tab:tab1} some numerical comparisons for the cases of $\Delta \ell=-2$, $\Delta \ell=0$, and $\Delta \ell=+2$.
A classic textbook case is of course the $3d \to 1s$ transition (this is a case where a sequence of E1 transitions, namely $3d \to 2p \to 1s$ will dominate
the overall transition strength). The semiclassical calculation based on re-scaling does not fare too well, being too small by a factor of about 3.
It is interesting to observe, however, that the agreement becomes quite good as the initial-state principal quantum number $n$ is increased.
Moving to higher initial values of $\ell$ shows that the semiclassical treatment yields respectable results.

For the $\Delta \ell= 0$ transitions we observe that for $\ell=\ell' =1$, i.e., p-state to p-state transitions, factor-of-two level agreement
is attained, and that the situation improves considerably as one moves to d-states, and higher angular momentum,
making the semiclassical result eqs.~(\ref{eq:rateSCL}-\ref{eq:ratedlpm}) a practical contender for evaluating transition
strengths, which could be useful for estimates of absorption of electromagnetic radiation.

\begin{table}[h]
  \centering
  \begin{tabular}{| l c c | l c c |}
  \hline
$(n, \ell) \to (n', \ell')$ & QM & SCL & $(n, \ell) \to (n', \ell')$ & QM & SCL \\
  \hline
 $3d \to 1s$ & 594. & 201. & $4f \to 2p$  & 61.8  & 70.1 \\
 $4d \to 1s$ & 327. & 256. & $4f \to 3p$  & 5.79  & 6.42 \\
 $5d \to 1s$ & 185. & 167. & $5g \to 3d$  & 11.6  & 13.4 \\
 $6d \to 1s$ & 112. & 107. & $5g \to 4d$  & 1.00  & 0.85 \\
  \hline
 $3p \to 2p$ & 23.9 & 14.7 & $4d \to 3d$  & 1.19  & 1.06\\
 $4p \to 2p$ & 10.3 & 6.29 & $5d \to 3d$  & 0.574  & 0.521 \\
 $5p \to 2p$ & 5.27 & 3.22 & $5f \to 4f$  & 0.153  & 0.150 \\
 $6p \to 2p$ & 3.05 & 1.86 & $6f \to 5f$  & 0.053  & 0.050 \\
  \hline
  $5p \to 4f$ & 0.047 & 0.158 & $6d \to 5g$  & 5.2(-3)  & 2.1(-2) \\
  $6p \to 4f$ & 0.032 & 0.124 & $7f \to 6h$  & 9.0(-4)  & 4.0(-3) \\
  $7p \to 4f$ & 0.021 & 0.089 & $8g \to 7i$  & 2.0(-4)  & 9.5(-4) \\
  $8p \to 4f$ & 0.014 & 0.064 & $9h \to 8k$  & 5.5(-5)  & 2.7(-4) \\
  \hline
  \end{tabular}
  \caption{Some quantum versus semiclassical E2 transition rates in decays per second. The column labeled QM shows quantum rates obtained from eq.~(\ref{eq:Rnlfin}),
  while the column labeled SCL shows the semiclassical rates according to eqs.~(\ref{eq:rateSCL}-\ref{eq:ratedlpm}). The top third of the table shows $\Delta \ell=-2$ transition
  rates, while the middle is for $\Delta \ell=0$, and the bottom third contains results for $\Delta \ell=+2$. The numbers in the bottom right part of the table are in a
  notation where $x.y(-i)$ represents $x.y \times10^{-i}$. 
  }
  \label{tab:tab1}
\end{table}

The case of $\Delta \ell= +2$, however, is somewhat disappointing, in that one finds at best order-of-magnitude agreement.
When comparing the semiclassical and quantum calculations we note that there are two aspects to consider.
On the one hand there is the radial matrix element of $r^2$, which enters equally for all three cases, $\Delta \ell = 0, \pm 2$.
The Fourier series of the elliptic orbit somehow encodes the right physics into the coefficients, such that a correspondence
with the squared quantum matrix element becomes possible. How do the calculations separate the transition strength into the parts that
correspond to $\Delta \ell = 0, \pm 2$ transtions respectively? This is accomplished in the semiclassical case by forming in eqs.~(\ref{eq:rateSCL}-\ref{eq:ratedlpm})
particular linear combinations of the coefficients $ A_k,  B_k,  C_k$.
In the quantum calculation, on the other hand, it is angular momentum coupling that dictates how the transition strength splits up.
The common aspect of these calculations is, of course, that the total angular momentum is conserved when taking into account the emitted
electromagnetic radiation. 
While we have no detailed understanding as to why the orbital angular momentum decreasing transitions
are described more favorably in the semiclassical approach, we note the similarity with the E1 dipole case~\cite{PhysRevA.71.020501},
where the semiclassical $\Delta \ell = -1$ branching fractions agreed much better with the quantum ones than those for $\Delta \ell = +1$.

\section{Conclusions}
\label{sec:conclusions}
This work expands on previous work for electric dipole transitions~\cite{PhysRevA.71.020501} in hydrogenic atoms demonstrating
that quantum-classical correspondence can be observed in a more complicated process connected with emission and absorption 
of electromagnetic waves with higher angular momentum. The semiclassical calculation involves a combination of mechanics
(description of the time evolution of the Kepler orbit by a Fourier series), and its coupling to electromagnetic radiation. The main
basis for evaluating the coupling of the orbital motion and the properties of the emitted radiation 
is provided in the graduate text Ref.~\cite{Jackson}. The quantum calculation was done mostly without sophistication by
brute force and computer algebra, but final expressions required the use of the Wigner-Eckart theorem in general form, as 
presented in the graduate text Ref.~\cite{Weissbluth}. While we provide the relevant expression for LSJ coupling we leave it
as an exercise for students to perform the comparison with the relativistic rates~\cite{Bunge}, which are available for higher $Z$,
and based on eq.~(\ref{eq:Rnljfin}) in the present work. Observation of what happens as $Z$ is increased would form a useful
exercise for graduate students. Finally, we may consider the present semiclassical calculation a useful warm-up exercise before
attempting a study of gravitational radiation in a post-Newtonian framework.

\begin{acknowledgments}
Financial support from the Natural Sciences and Engineering Research Council of Canada (NSERC) is acknowledged. 
Marko Horbatsch acknowledges contributions made by Nick Hamer during the early stages of this project, 
and his support through an NSERC undergraduate research award.
\end{acknowledgments}

%

\bibliography{QuadrupoleRadiation}

\end{document}